\documentclass{elsart}
\journal{Physics Letter A}

\usepackage{graphicx,subfigure}

\usepackage{amssymb,amsmath,bm}
\usepackage{cite}
\usepackage{url}

\newcommand\mymatrix[1]{\bm{\mathrm{#1}}}

\begin{document}

\newlength\figwidth
\setlength\figwidth{0.5\columnwidth}
\newlength\imgwidth
\setlength\imgwidth{0.3\columnwidth}

\begin{frontmatter}
\title{Cryptanalysis of an image encryption scheme based on a new total shuffling algorithm}
\author[Spain]{David Arroyo\corauthref{corr}},
\author[hk-cityu]{Chengqing Li},
\author[germany]{Shujun Li},
\author[Spain]{Gonzalo Alvarez} and
\author[germany]{Wolfgang A. Halang}
\corauth[corr]{Corresponding author: David Arroyo
(david.arroyo@iec.csic.es).}
\address[Spain]{Instituto de F\'{\i}sica Aplicada, Consejo Superior de
Investigaciones Cient\'{\i}ficas, Serrano 144, 28006 Madrid, Spain}
\address[hk-cityu]{Department of Electronic Engineering, City University of Hong Kong,
83 Tat Chee Avenue, Kowloon Tong, Hong Kong SAR, China}
\address[germany]{FernUniversit\"{a}t in Hagen, Chair of Computer Engineering, Universit\"{a}tsstra{\ss}e 27, 58084 Hagen, Germany}

\begin{abstract}
Chaotic systems have been broadly exploited through the last two
decades to build encryption methods. Recently, two new image
encryption schemes have been proposed, where the encryption
process involves a permutation operation and an XOR-like
transformation of the shuffled pixels, which are controlled by
three chaotic systems. This paper discusses some defects of the
schemes and how to break them with a chosen-plaintext attack.
\begin{keyword}
Chaotic encryption, Lorenz system, Chen's system, hyper-chaos,
logistic map, chosen-plaintext attack, permutation-only encryption
algorithms, cryptanalysis \PACS 05.45.Ac, 47.20.Ky.
\end{keyword}
\end{abstract}
\end{frontmatter}

\section{Introduction}

When we think about exchanging information we are very interested in
finding a way to make it fast and secure. Modern telecommunications
technologies allow to send and receive files, images, and data in a
relatively short time depending on the bandwidth available.
Nowadays, the use of traditional symmetric and asymmetric
cryptography is the way to secure the information exchange
\cite{menezes97, schneier96}. However, applications involving
digital images and videos demand other encryption schemes. Indeed,
the bulky size and the large redundancy of uncompressed
videos/images make it necessary to look for new methods to deal with
those features in order to facilitate the integration of the
encryption in the whole processing procedure. For recent surveys on
image and video encryption, please refer to
\cite{shujun:bookMultimedia,book:DRM,book:MultimediaEncryption,book:MultimediaEncryption2}.

The main features of chaotic systems (sensitivity to initial
conditions, ergodicity, mixing property, simple analytic
description and high complex behavior) make them very interesting
to design new cryptosystems. Image encryption is an area where
chaos has been broadly exploited. In fact, chaotic systems have
been used to mask plain-images through XOR-like substitution
operations \cite{imageEncryption:hunChen03}, spatial permutation
\cite{Yen&Guo:ChaoticImageEncryption:IEEPVISP2000} or the
combination of both techniques \cite{imageEncryption:chen04}. This
paper is focused on two image encryption schemes proposed in
\cite{gao07a,gao07b}. In both papers the image encryption is based
on a secret permutation derived from the logistic map, and a
masking of the gray-scale values of the shuffled pixels with a
keystream generated from one or two chaotic systems. The only
difference between the two encryption schemes is that in
\cite{gao07a} two chaotic systems (Lorenz and Chen's systems) are
used to generate the keystream, while in \cite{gao07b} only one
hyper-chaotic system is used. Because such a difference is
independent of the security, we only focus on the cryptanalysis of
the scheme proposed in \cite{gao07a}.

The rest of this paper is organized as follows. The scheme under
study is described briefly in the next section. In Sec.
\ref{sec:defects} some important problems of the cryptosystem are
remarked. Then, a chosen-plaintext attack is described in
Sec.~\ref{sec:cryptanalysis} along with some experimental results.
In the last section the conclusion is given.

\section{The encryption scheme}
\label{sec:EncryptionScheme}

Assuming that the size of the plain-image $\mymatrix{I}$ is
$M\times N$ and the cipher-image is $\mymatrix{I}'$, the
encryption scheme proposed in \cite{gao07a} can be described by
the following two procedures. Please note that we use different
notations from the original ones in \cite{gao07a} to get a simpler
and clearer description.

\begin{itemize}
\item \textit{Shuffling procedure}

In this procedure, the plain-image $\mymatrix{I}$ is permuted to
form an intermediate image $\mymatrix{I}^*$ according to a total
shuffling matrix $\mymatrix{P}^*$, which is derived by
pseudo-randomly permuting the rows and columns of the original
position matrix $\mymatrix{P}=[(i,j)]$. The pseudo-random row and
column permutations are generated by iterating the logistic map
$x_{n+1}=4x_n(1-x_n)$ from a given initial condition $x_0$.

\item \textit{Masking procedure}

In this procedure, the intermediate image $\mymatrix{I}^*$ is
further masked by a keystream $\{B(i)\}_{i=1}^{MN}$ as follows:
$\forall i=1\sim MN$, $I'(i)=I^*(i)\oplus B(i)\oplus I'(i-1)$, where
$I(i)$, $I'(i)$ denote the $i$-th pixels of $\mymatrix{I}^*$ and
$\mymatrix{I}'$ (counted from left to right and from top to bottom),
respectively, and $I'(0)=128$.

The keystream $\{B(i)\}_{i=1}^{MN}$ is generated by iterating the
Lorenz and Chen's systems and doing some postprocessing on all the
6 chaotic variables (the first $N_0$ iterations of Lorenz system
and the first $M_0$ iterations of Chen's systems are discarded to
enhance the security). Because our cryptanalysis succeeds
regardless of the keystream's generation process, we ignore this
part and readers are referred to Sec.~2.3 of \cite{gao07a} for
details.
\end{itemize}

In \cite{gao07a}, it is claimed that the secret key includes the
initial values of the Lorenz and Chen's systems and the number of
initial iterations $N_0$, $M_0$. It is quite strange why the
initial condition of the logistic map is not claimed to be part of
the key, since the image encryption scheme is based on ``a new
total shuffling algorithm'' (as can be seen in the title of
\cite{gao07a}). In this cryptanalysis paper, we assume that the
initial condition of the logistic map is also part of the key. We
believe it is also the original intention of the authors of
\cite{gao07a}. In addition, note that both $\mymatrix{P}^*$ and
$\{B(i)\}_{i=1}^{MN}$ are independent of the plaintext and
ciphertext, so they can be used as an equivalent key.

\section{Design weaknesses}
\label{sec:defects}

In this section, we discuss some defects of the scheme under
study.

\subsection{Low sensitivity to the change of plain-image}
\label{Sec:Insensitivity}

\begin{figure}
\centerline{%
\subfigure[$$]{\includegraphics{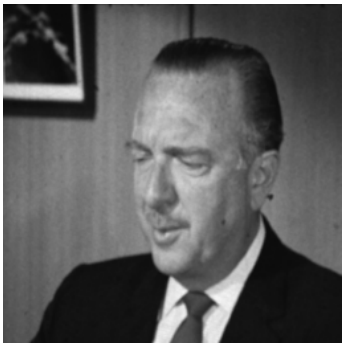}}
\subfigure[$$]{\includegraphics{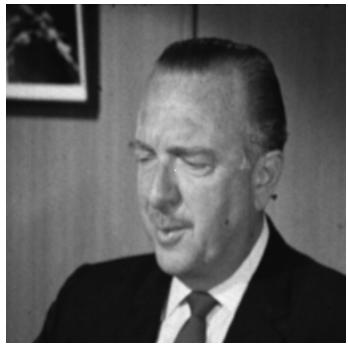}}%
}
\centerline{%
\subfigure[$$]{\includegraphics{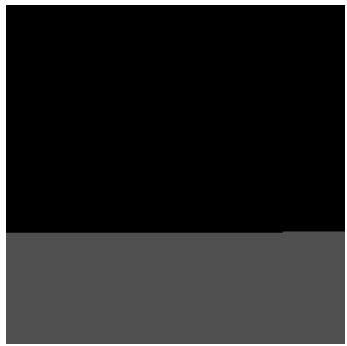}}%
}%
\caption{Illustration of the low sensitivity to the change of the
plain-image: (a) the first plain-image $\mymatrix{I}_0$; (b) the
second plain-image $\mymatrix{I}_1$ (only the center pixel is
different from $\mymatrix{I}_0$); (c) the differential
cipher-image
$\mymatrix{I}_0'\oplus \mymatrix{I}_1'$.}%
\label{fig:lowSensitivity}
\end{figure}

It is well known that the ciphertext of a secure encryption scheme
should be very sensitive to the change of plaintext \cite[Rule
9]{Alvarez06a}. Unfortunately, the encryption scheme under study
fails to satisfy this requirement. Given two plain-images
$\mymatrix{I}_0$ and $\mymatrix{I}_1$ with only one pixel
difference at the position $(i,j)$, the difference will be
permuted to a new position $(i^*,j^*)$ according to the shuffling
matrix $\mymatrix{P}^*$. Then, because all plain-pixels before
$(i^*,j^*)$ are identical for the two plain-images, the
ciphertexts will also be identical. This shows the low sensitivity
of the image encryption scheme to changes in the plain-image.
Figure~\ref{fig:lowSensitivity} gives an example of this problem.
It can be seen how the differential cipher-image is equal to zero
for any pixel before $(i^*,j^*)$ and equal to a constant value
after that position.

\subsection{Reduced Key space}

As claimed in \cite{gao07a}, $N_0$ and $M_0$ are also part of the
key. However, from an attacker's point of view, he/she only needs to
guess the chaotic states after the $N_0$ and $M_0$ chaotic
iterations as the initial conditions of the Lorenz and Chen's
systems. In this way, $N_0$ and $M_0$ are removed from the key and
the key space is reduced.

\subsection{Problem with chaotic iterations of Lorenz and Chen's systems}

In \cite{gao07a}, the authors did not say anything about the time
step $\tau$ of iterating the Lorenz and Chen's systems. However,
the randomness of the keystream $\{B(i)\}_{i=1}^{MN}$ is tightly
dependent on the value of time step. As an extreme example, if
$\tau=10^{-20}$, we will get a keystream of identical elements
(according to the algorithm described in Sec.~2.3 of
\cite{gao07a}). As a matter of fact, the value of $\tau$ is
dependent on the multiplication factor $10^{13}$ occurring in
Step~4 of the encryption process (see Sec.~2.3 of \cite{gao07a}):
$x_i=\bmod((\mathrm{abs}(x_i)-\mathrm{Floor}(\mathrm{abs}(x_i)))\times
10^{13},256)$.

\subsection{Low encryption speed}

Because the chaotic iterations of Lorenz and Chen's systems involve
complicated numerical differential functions, the encryption speed
is expected to be very slow compared with other traditional ciphers.
To asses this fact, we derived a modified encryption scheme from the
original one by replacing the Lorenz and Chen's systems with the
logistic map, and then compared the encryption speeds of the two
cryptosystems. Both cryptosystems were implemented using MATLAB on a
PC with a 1.6GHz processor and 512MB of RAM. For images of size
$256\times 256$, the typical encryption time for the original
cryptosystem in \cite{gao07a} was around 5.8 seconds, while the
modified cryptosystem based on the logistic map required in average
around 1.2 seconds to encrypt an image. The experiments have clearly
shown that using continuous chaotic systems can drastically reduce
the encryption speed. Since there are also no other obvious merits
in using continuous chaotic systems rather than a simple
discrete-time chaotic map, the use of the Lorenz and Chen's systems
in the image encryption scheme under study is unnecessary. Instead,
these continuous chaotic systems can be replaced by a simpler
discrete-time chaotic map without compromising the security.

\section{Chosen-plaintext attack}
\label{sec:cryptanalysis}

When a variation of stream cipher is created, as in the case under
study, obtaining the keystream is totally equivalent to obtaining
the key whenever different plain-images are encrypted using the
same key. In this section, we present a chosen-plaintext attack
which allows to recover both the keystream and the shuffling
matrix.

Let us choose a plain-image $\mymatrix{I}_1$ such that $\forall
i,j=1\sim MN$, $I_1(i)=I_1(j)=a$. In this case, the shuffling part
does not work, so we have $\mymatrix{I}^*_1=\mymatrix{I}_1$. Then,
we can recover the keystream as follows: $\forall i=1\sim MN$,
$B(i)=I_1(i)\oplus I'_1(i)\oplus I'_1(i-1)$.

After removing the masking part, we can try to recover the
shuffling matrix. According to the general cryptanalysis on
permutation-only ciphers in \cite{shujunLi04a}, only
$\lceil\log_{256}(MN)\rceil$ chosen plain-images are needed to
recover the shuffling matrix $\mymatrix{P}^*$. In total we need
$\lceil\log_{256}(MN)\rceil+1$ chosen plain-images to perform this
chosen-plaintext attack.

\begin{figure}[!htb]
\center
\begin{minipage}{0.8\figwidth}
\centering
\includegraphics[width=\textwidth]{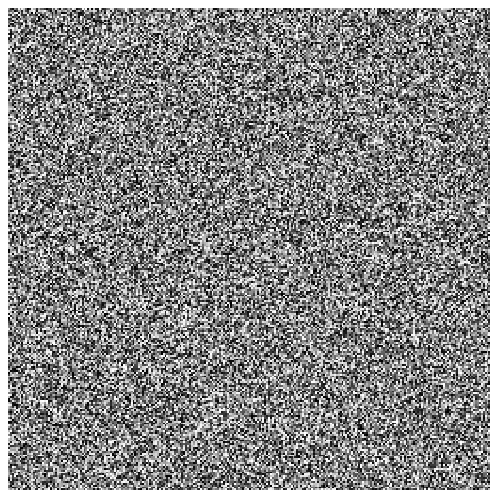}
(a)
\end{minipage}
\begin{minipage}{0.8\figwidth}
\centering
\includegraphics[width=\textwidth]{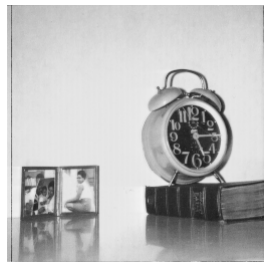}
(b)
\end{minipage}
\caption{The result of the chosen-plaintext attack: (a) a
cipher-image encrypted with the key as shown in
Table~\ref{table:key}; (b) the decrypted plain-image using the
equivalent key $\left(\mymatrix{P}^*,\{B(i)\}_{i=1}^{MN}\right)$
obtained via the chosen-plaintext attack.} \label{experiment}
\end{figure}

With the aim of verifying the proposed attack, several experiments
have been done. One of the examples is shown in
Fig.~\ref{experiment}, where the images are of size $256\times
256$ and the secret key involved is shown in
Table~\ref{table:key}. As it was mentioned above, the shuffling
process is broken using $\log_{256}(MN)=2$ chosen plain-images,
while the masking procedure cryptanalysis requires one chosen
plain-image. The three chosen plain-images allow to decipher the
cipher-image included in Fig.~\ref{experiment}(a) and thus to get
the corresponding plain-image (Fig.~\ref{experiment}(b)), even
when the secret key is unknown.

\begin{table}[!h]
\caption{Key value used in the experiment.}
\begin{tabular}{c|c|c|c|c|c|c|c|c}
\hline
$x_1(0)$&$x_2(0)$&$x_3(0)$&$x_4(0)$&$x_5(0)$&$x_6(0)$&$N_0$&$M_0$&$x_0$\\\hline
$0.3$& $-0.4$&$ 1.2$&$ 10.2$&$
-3.5$&$4.4$&$3000$&$2000$&$0.4$\\\hline
\end{tabular}
\label{table:key}
\end{table}

\section{Conclusions}

The security of the image encryption scheme proposed in
\cite{gao07a} has been analyzed in detail. The cryptanalytic
results are also valid for the other scheme proposed in
\cite{gao07b}. It has been shown that the equivalent secret key
can be recovered in a chosen-plaintext attack with only
$\lceil\log_{256}(MN)\rceil+1$ chosen plain-images. In addition,
some other defects have also been distinguished in the scheme
under study. Among those defects, it is necessary to emphasize the
one concerning the encryption speed, since it informs about the
non-convenience of continuous-time chaotic systems for
implementing fast encryption procedures. The weak security
properties frustrate the usage of the scheme in practice.

\section*{Acknowledgments}

The work described is this letter was partially supported by
Minis\-terio de Educaci\'on y Ciencia of Spain, Research Grant
SEG2004-02418. Shujun Li was supported by the Alexander von Humboldt
Foundation, Germany.

\bibliographystyle{elsart-num}
\bibliography{database}

\end{document}